# Understanding the flat thermal conductivity of $La_2Zr_2O_7$ at ultrahigh temperatures

Hao Zhou, Janak Tiwari, Tianli Feng*

Department of Mechanical Engineering, University of Utah, Salt Lake City, Utah 84112, USA

**Corresponding Authors**

*Email: tianli.feng@utah.edu

**Abstract**

Many crystals, such as lanthanum zirconate ($La_2Zr_2O_7$), exhibit a flat temperature dependence of thermal conductivity at elevated temperatures. This phenomenon has recently been attributed to the inter-band phonon tunneling (or diffuson) contribution using different formalisms. However, the contributions of finite-temperature corrections (e.g., higher-order phonon-scattering, phonon renormalization, and phonon scattering cross-section softening effects) remain unclear. In this work, we predict and compare the thermal conductivity of $La_2Zr_2O_7$ using three distinct first-principles methods. The first method is Green-Kubo molecular dynamics (MD) based on temperature-dependent machine learning interatomic potentials (MLIPs) trained from ab initio MD simulations, which successfully predict the flat trend at ultra-high temperatures. The second method is the Peierls Boltzmann transport equation (BTE), within the phonon particle framework, using phonon lifetime that includes all the finite-temperature corrections. Four-phonon scattering is found large but is cancelled by the phonon scattering cross-section softening effect. As a result, BTE with temperature corrections does not reproduce the flat thermal conductivity. The third method is Wigner formalism, which includes both phonon particle and wave contributions, which successfully reproduce the flat thermal conductivity. Diffuson and phonon contribute about 67% and 27% of thermal conductivity at 1800 K, respectively. The radiation contribution to thermal conductivity is around 6%. The scaling laws of the phonon, diffuson, radiation, and total thermal conductivity are found to be $\sim T^{-0.97}$, $\sim T^{0.43}$, $\sim T^{2.01}$, and $\sim T^{-0.40}$, respectively. This work clarifies the thermal transport mechanisms in $La_2Zr_2O_7$ at ultra-high temperatures from different aspects.

## I. Introduction

Thermal barrier coatings are vital for many cutting-edge technologies that require high temperatures environment [1] such as nuclear power plants [2], hypersonic vehicles [3], thermal energy storage [4], advanced turbines [5], etc. Among the emerging thermal barrier coating materials, lanthanum zirconate ($La_2Zr_2O_7$) stands out due to its exceptional properties, including ultrahigh melting point, physical stability, thermal compatibility, and ultralow thermal conductivity ($\kappa$) [6,7]. To utilize and engineer $La_2Zr_2O_7$ based composites, it is necessary to understand its intrinsic thermal transport mechanisms at ultra-high temperatures. This may also shed light on the properties of many other $A_2B_2O_7$ type of thermal barrier coating materials.

Heat in crystals is primarily carried by the quanta of lattice vibration, namely, phonons. Thermal resistance is a result of phonon scattering, which includes three-phonon scattering (3ph, the lowest order scattering mechanism) and four-phonon scattering (4ph, a higher order mechanism). The 3ph rate, $\tau_3^{-1} \sim n \sim T$, increases linearly with phonon population ($n$) and temperature ($T$) at elevated temperatures, resulting in the commonly seen $\kappa \sim T^{-1}$ decay trend for crystals. Recently, it has been recognized that 4ph is important at high temperatures since $\tau_4^{-1} \sim n^2 \sim T^2$ increases quadratically with phonon population and temperature. With 4ph, the thermal conductivity of crystals under the phonon particle framework can decay faster than $T^{-1}$, i.e., with $\kappa \sim T^{-\alpha > 1}$ at high temperatures. However, the thermal conductivity of La$_2$Zr$_2$O$_7$ shows a flat temperature dependence of thermal conductivity at elevated temperatures [6,8–12], which challenges the phonon particle and Peierls Boltzmann transport equation (BTE) frameworks.

The high-temperature flat thermal conductivity of complex crystals has very recently been interpreted using the diffuson picture, emphasizing inter-band phonon tunneling beyond the conventional phonon particle (or Peierls BTE) framework. Specifically, it has been argued that as temperature rises, phonons may become ill-defined when their lifetime ($\tau$) becomes smaller than period ($P$), or mean free path (MFP) becomes shorter than wavelength or than the minimum atomic distance [13]. In this scenario, phonons cannot be treated as particles only, and their wavelike nature allows them to tunnel between close eigenstates. The diffuson picture has been implemented into many theoretical models to explain the flat thermal conductivity for many materials [14–18]. A most recent and rigorous theory that can be integrated with first-principles calculations is the Wigner formalism [19,20], developed by Simoncelli *et al*. Based on the Wigner formalism, for complex crystals that have small inter-band spacing or large phonon linewidth broadening, diffuson can dominate heat transport, leading to a glass-like temperature dependence of thermal conductivity at elevated temperatures.

Apart from the diffuson theory, the finite-temperature corrections within the phonon particle (or Peierls BTE) framework were also found to cause the flat thermal conductivity for some materials at high temperatures [21–24]. The aforementioned decreasing trend of $\kappa \sim T^{-\alpha>1}$ is built on the assumption that all the phonon properties, except for phonon population, including phonon dispersion and scattering cross-section (probability), are based on the lattice structure and interatomic force constants (FCs) at 0 K or ground state (GS). If the temperature dependent (TD) effects, including the finite-temperature corrections to the lattice constant as well as the second-, third-, and fourth-order interatomic force constants, are considered, the $\kappa \sim T^{-\alpha}$ trend would be changed and flattened. Specifically, the correction to the lattice constant and 2nd-FCs can change the phonon dispersion and then decrease three-phonon scattering phase spaces [25]. The correction to the 3rd- and 4th-FCs can decrease the phonon scattering cross sections [26]. All these factors tend to decrease the phonon scattering rate, compensating for the increasing trend caused by the increased phonon population with temperature. As a result, the thermal conductivity would not decrease as fast as $\kappa \sim T^{-\alpha>1}$. This effect is material dependent. In the case of $La_2Zr_2O_7$, while the diffuson picture has been employed to elucidate the flat trend in thermal conductivity at ultra-high temperatures [17,20], the role of 4ph and finite-temperature corrections remains unclear. These factors as mentioned above can affect the temperature trend at ultra-high temperatures significantly. Additionally, determining the extent to which diffuson contributes to the observed flat trend at ultra-high temperatures requires further investigation.

In this work, we predict and compare the thermal conductivity of $La_2Zr_2O_7$ using three distinct first-principles methods. The first method is Green-Kubo molecular dynamics (MD) based on temperature-dependent machine learning interatomic potentials (MLIPs) [27] trained from ab initio MD simulations [28,29]. GKMD effectively captures all finite-temperature corrections and all orders of anharmonicities. Thermal expansion is considered in MD simulations. The second method is the Peierls Boltzmann transport equation (BTE), within the phonon particle framework, including all temperature corrections. Since the calculations of temperature-dependent fourth-order force constants and four-phonon scattering are extremely time consuming, we employ phonon spectral energy density (SED) analysis based on MD trajectories to extract phonon lifetime. The SED method can naturally include all finite-temperature corrections and all orders of anharmonicities. The third method is the Wigner formalism. In addition, radiation contribution to thermal conductivity is rigorously calculated to clarify the long-standing question about its contribution to the flat/increasing trend of experimental data.

The remainder of this paper is organized as follows. In Sec. II, we explain the methodology, which includes Sec. II.A, quasi-harmonic approximation (QHA) for thermal expansion coefficient (TEC) calculation, Sec. II.B, the training and testing processes of MLIP, Sec. II.C, thermal conductivity calculations by GKMD simulations, Sec. II.D, thermal conductivity calculations by BTE with lifetime obtained from SED, and Sec. II.E, thermal conductivity calculations by the Wigner formalism. In Sec. III, we present the TEC results (Sec. III.A), MLIP training and testing (Sec. III.B), thermal conductivity obtained from GKMD (Sec. III.C), phonon lifetime and thermal conductivity by SED-BTE (Sec. III.D), phonon and diffuson thermal conductivity by Wigner formalism (Sec. III.E), and photon contribution to thermal conductivity (Sec. III.F). The conclusions are summarized in Sec. IV.

**II. Methodology**

The methodology workflow of this work is summarized in Fig. 1. First, the primitive cell of $La_2Zr_2O_7$ is relaxed at the ground state. Then, the thermal expansion coefficient (TEC) is calculated within the framework of quasi-harmonic approximation (QHA). With TEC applied, *ab initio* molecular dynamics (AIMD) simulations are performed at various temperatures to construct training database for machine learning interatomic potential. With the obtained machine learning interatomic potential, we first run molecular dynamics (MD) and utilize Green-Kubo theory to obtain thermal conductivity, $\kappa_{GK}$. Based on molecular dynamics trajectories, mode-dependent phonon lifetimes are extracted by using phonon spectral energy density (SED) analysis at different temperatures. These phonon lifetimes are used to calculate thermal conductivity within phonon particle and Peierls BTE frameworks, i.e., $\kappa_{SED}$. Apart from this, second-order, third-order, and fourth-order interatomic force constants are computed using finite difference method based on the relaxed structure. The Wigner formalism is used to calculate the phonon particle and diffuson wavelike contribution, namely. $\kappa_{ph}$ and $\kappa_D$, respectively, using the phonon lifetime obtained based on the FCs. Last but not least, we rigorously calculate the radiation contribution to thermal conductivity using the Lorentz oscillator model combined with the Rosseland model.

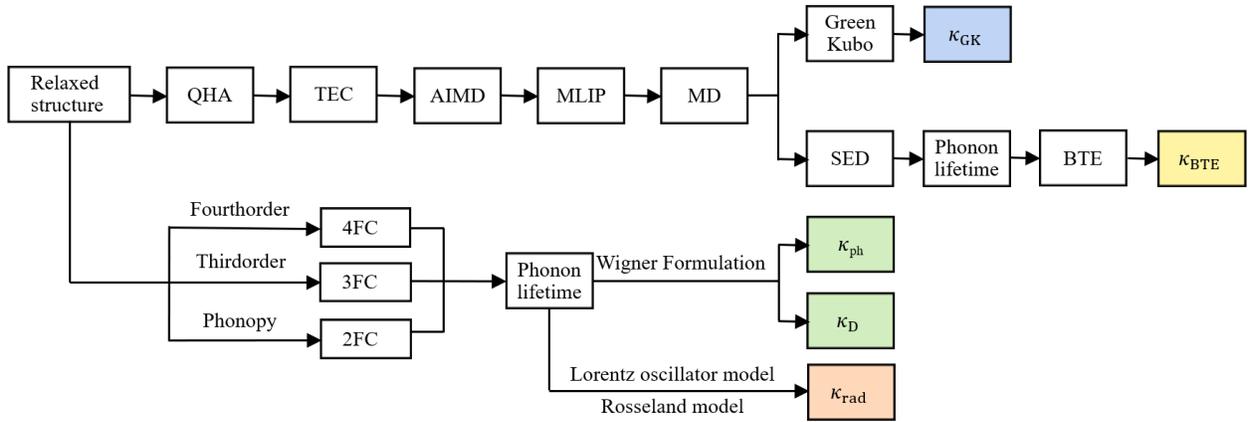

FIG. 1. Methodology employed in this work.

## A. Thermal expansion coefficient

To account for the effect of thermal expansion, the linear TEC $\alpha_L$ of La$_2$Zr$_2$O$_7$ is calculated using the QHA as [30,31]

$$\alpha_L(T) \equiv \frac{1}{a(T)} \frac{da(T)}{dT} = -\frac{k_B}{3N_\mathbf{q} V_c B} \sum_\lambda \gamma_\lambda \cdot \left(\frac{x}{2}\right)^2 \cdot \left[1 - \coth^2\left(\frac{x}{2}\right)\right], \tag{1}$$

where $a$ is lattice constant as a function of temperature $T$, $k_B$ is the Boltzmann constant, $N_\mathbf{q}$ is the number of $\mathbf{q}$ points, $V_c$ is the volume of the primitive cell, $x$ stands for $\hbar\omega/k_B T$, $\hbar$ is the reduced Planck's constant, and $\omega$ is the phonon angular frequency. $\lambda \equiv (\mathbf{q}, j)$ is a shorthand notation for a phonon mode with wavevector $\mathbf{q}$ and polarization branch $j$. The summation goes over all the $3N_q n_b$ phonon modes, where $n_b = 22$, the number of basis atoms in each primitive cell. $B$ is the bulk modulus and is calculated by $B = -VdP/dV$, which measures the ability to withstand changes in volume $V$ with pressure $P$. In this work, $B = -VdP/dV$ is calculated by measuring the pressure change of simulation domain at different volumes at the ground state. $\gamma_\lambda$ is the mode-dependent Grüneisen parameter, which can be obtained via two approaches. One is quasi-harmonic approximation (QHA), which is to calculate the phonon dispersion at different cell volumes and then measure the relative frequency change with the relative cell volume change, given by its definition $\gamma_\lambda = -V/\omega \cdot \partial\omega/\partial V$. The other approach is to use the third-order anharmonic force constants [32,33], ignoring the higher-order anharmonic force constants. The obtained linear TEC is applied to the following GKMD and SED calculations at each temperature. Based on Eq. (1), the temperature-dependent lattice constant $a(T)$ can be obtained by

$$a(T) = a_{0K} \exp \int_0^T \alpha_L(T) \, dT \tag{2}$$

Note that the $\alpha_L(T)$ represents the instant lattice constant change around the given temperature $T$. This is different from some reports [34], where the thermal expansion coefficient, labeled as $\alpha_L^*(T)$ here, is defined as the lattice constant change relative to a fixed temperature, often room temperature: $\alpha_L^*(T) \equiv \frac{a(T)-a_{300K}}{a_{300K}\cdot(T-300K)} \neq \alpha_L(T)$.

All the calculations are performed within the density functional theory (DFT) framework using Vienna ab-initio simulation package (VASP) [35]. The calculations employ the projector augmented wave (PAW) [36] method and the PBEsol exchange-correlation functional [37,38] with a plane-wave energy cutoff of 500 eV. The ground-state lattice constants and atomic positions are relaxed with a 6×6×6 **k**-mesh, an electron energy convergence threshold of $10^{-8}$ eV, and a force convergence threshold of $10^{-5}$ eV/Å. The second-order force constants and phonon dispersion are calculated by Phonopy [39]. The Grüneisen parameter calculation from the third-order anharmonic force constants are conducted by using ShengBTE [32,33].

### B. Moment tensor potential

In the past decade, several types of machine-learning interatomic potentials were developed to fit the *ab initio* energy potential surface, which were found to have higher accuracy than empirical analytical formula-based potentials [40–42]. Among them, the MTP developed by Novikov *et al.* using the MLIP package [27] has been proven to exhibit small error and low computational cost simultaneously [43]. Therefore, we take advantage of MTP to describe the interatomic interactions in $La_2Zr_2O_7$. The training database is prepared by *ab initio* molecular dynamics (AIMD) using VASP [35] with canonical ensemble (NVT) and time step of 2 fs. The plane-wave energy cutoff is 500 eV with $10^{-7}$ eV electron energy convergence threshold and a **Γ** point only **k**-mesh to accelerate the simulation. A supercell of 2×2×2 primitive cells containing 176 atoms is adopted in the simulation domain. The temperatures used in AIMD are 300 K, 1200 K, 1500 K, and 1800 K, with TEC applied to the lattice constant correspondingly. Four independent AIMDs with randomly displaced initial atomic positions are performed for each temperature to better sample the potential energy surface. Each simulation is 1.3 ps long. Atomic positions, energies, forces, and stresses are recorded to construct the training (19,548 snapshots) and testing (502 snapshots) database. An MTP is trained to account for the atomic interactions of temperatures ranging from 300 K to 1800 K. Before the training process, the potential level test is conducted. A higher-level potential contains more fitting parameters and requires a larger computational load. After selecting an appropriate level, we train the MTP with 1000 iteration steps with the minimum and maximum atomic interaction cutoffs of 1.2 Å and 5.0 Å, respectively. The minimum cutoff is smaller than the minimum atomic distance in both training and testing database. The maximum cutoff is large enough to include the dominant interatomic interactions, and smaller than half of the AIMD simulation dimension so that it can be applied to any extended structures. Root mean square error (RMSE) is selected to evaluate the potential quality.

### C. Green-Kubo molecular dynamics

Once the MTP with a low error is developed, GKMD is performed by using LAMMPS package [44] to obtain the lattice thermal conductivity. A 4×4×4 supercell of conventional cell containing 5632 atoms is adopted. The size effect is studied systematically as shown in Sec. III. The time step of GKMD is 1 fs, and periodic boundary conditions are implemented in all three directions. The simulation is carried out first by 0.2 ns NVT, a 0.2 ns NVE to stabilize the system, and another 2 ns NVE to calculate the heat current autocorrelation function. The correlation time is set as 50 ps, which is long enough to capture the fluctuation dissipation since most phonon modes have a lifetime much shorter than this. Based on the Kubo formula [28,29], $\kappa_{GK}$ is obtained as

$$\kappa_{GK} = \frac{1}{3k_B T^2 V} \int_0^\infty \langle \vec{J}(0) \cdot \vec{J}(t) \rangle dt, \tag{3}$$

where $V$ is the volume of the total simulation domain, $T$ is temperature, $\vec{J}(t)$ is the heat current, and the angular bracket represents an autocorrelation. To eliminate the intrinsic statistical error of GKMD, four independent runs with different initial velocities are conducted and averaged. The ratio between the total simulation time of the last NVE and the correlation time is larger than 300, which is found sufficient [45]. The error bar is presented by calculating standard error [45]. Note that the temperatures focused on in this study are relatively high, and the difference between classical and quantum statistics is negligible.

**D. Phonon spectral energy density analysis and thermal conductivity**

SED analysis can extract phonon lifetimes from MD trajectories that naturally incorporate all temperature effects and all orders of anharmonicity. SED is a Fourier transformation of velocity normal mode coordinates and can be fitted by the Lorentzian function for each phonon mode to get the linewidth and frequency [46,47]. The phonon SED $\Phi_s(\mathbf{q}, \omega)$ for a given $\mathbf{q}$ and branch $j$ is calculated as

$$\Phi_j(\mathbf{q}, \omega) = \left| \dot{q}_{\mathbf{q},j}(\omega) \right|^2 = \left| \int_0^{+\infty} \dot{q}_{\mathbf{q},j}(t) e^{-i\omega t} dt \right|^2, \tag{4}$$

where $\dot{q}_{\mathbf{q},j}(\omega)$ is the Fourier transformation of the time derivative of normal mode coordinates, $\dot{q}_{\mathbf{q},j}(t)$:

$$\dot{q}_{\mathbf{q},j}(t) = \sum_\alpha^3 \sum_b^{n_b} e_\alpha^{*b}(\mathbf{q}, j) \sum_l^{N_c} \sqrt{\frac{m_b}{N_c}} \dot{u}_\alpha^{l,b}(t) \exp(i\mathbf{q} \cdot \mathbf{r}_0^l) \tag{5}$$

where $e^*$ is the complex conjugate of the eigenvector component, $\dot{u}(t)$ is time-dependent atomic velocity, $r_0$ is the equilibrium position, $m$ is mass, $n_b$ is the number of basis atoms in one unit cell, $N_c$ is the number of unit cells in the simulation domain, $\alpha$ is Cartesian coordinate, $l$ is the index for unit cell, and $b$ is the index of basis atoms in a unit cell. The total SED ($\Phi(\mathbf{q},\omega)$) for a certain $\mathbf{q}$ point can be obtained by summing up the SED data of all the phonon branches ($\Phi(\mathbf{q},\omega) = \sum_j \Phi_j(\mathbf{q},\omega)$). After Fourier transformation, $\Phi_j(\mathbf{q},\omega)$ is smoothed by adjacent-averaging method and then fitted by Lorentzian function for every single peak as

$$\Phi_j(\mathbf{q},\omega) = \frac{C_{\mathbf{q},j}}{\left(\omega - \omega_{\mathbf{q},j}^A\right)^2 + \Gamma_{\mathbf{q},j}^2} \tag{6}$$

where $C_{\mathbf{q},j}$ is a mode-dependent constant, $\omega_{\mathbf{q},s}^A$ is the peak frequency position, and $\frac{1}{2\Gamma_{\mathbf{q},j}}$ is the phonon lifetime. Then, thermal conductivity ($\kappa_{\text{SED}}$) is calculated within the framework of the BTE using the relaxation time approximation (RTA) as

$$\kappa_{\text{SED}} = \sum_\lambda c_\lambda v_{\alpha,\lambda}^2 \tau_{\lambda,\text{SED}} \tag{7}$$

where $c$ is specific heat, $v_\alpha$ is group velocity along the $\alpha$ direction, $\tau_{\text{SED}}$ is the phonon lifetime obtained from SED analysis. For low thermal conductivity materials, the RTA should not yield a considerable error compared to the iteration exact solution to BTE.

To obtain MD trajectories, a 6×6×6 cubic supercell of conventional cell of La$_2$Zr$_2$O$_7$ is employed. The periodic boundary condition is applied to all three dimensions. We first perform a 400,000-step NPT, followed by a 400,000-step NVE to fully relax the structure. Then, a 4,000,000-step NVE is performed to record the velocity data. The time step used in this study is 1 fs. A 6×6×6 $\mathbf{q}$-mesh is employed to extract phonon lifetime and thermal conductivity.

## E. Thermal conductivity based on Wigner formalism

For comparison, we calculate the phonon and diffuson thermal conductivity by using the Wigner formalism [19,20]. Four-phonon scattering is included. The 2$^{nd}$- and 3$^{rd}$-FC are extracted via finite difference method (FDM) with a 3×3×3 **k**-mesh using Phonopy [48] and Thirdorder [33] packages, respectively. All the other settings of the self-consistent calculations are the same as those in lattice relaxation. Up to fifth nearest neighbors are considered in the extraction of 3$^{rd}$-FC. The three-phonon scattering rates $(1/\tau_{3,\lambda}^0)$ are calculated based on 3$^{rd}$-FC as [49]

$$\frac{1}{\tau_{3,\lambda}^0} = \sum_{\lambda'\lambda''} \left\{ \frac{1}{2}(1 + n_{\lambda'}^0 + n_{\lambda''}^0)\mathcal{L}_- + (n_{\lambda'}^0 - n_{\lambda''}^0)\mathcal{L}_+ \right\} \quad (8)$$

Here $\lambda' = (\mathbf{q}', j')$ and $\lambda'' = (\mathbf{q}'', j'')$ go over all the phonon modes except for $\lambda = (\mathbf{q}, j)$. $n_\lambda^0 = [\exp(\hbar\omega_\lambda/k_B T) - 1]^{-1}$ is the phonon Bose-Einstein distribution function. The formalisms of $\mathcal{L}_+$ and $\mathcal{L}_-$ can be found in Ref. [50,51]. The 4$^{th}$-FC are calculated via FDM using Fourthorder [52] package in this work. Due to the computational cost, the settings in 4$^{th}$-FC extraction are different from the former calculations. The plane-wave energy cutoff is 450 eV, and the electron energy convergence threshold is 10$^{-7}$ eV. Only the first nearest neighbor is considered in the extraction of 4$^{th}$-FC. The four-phonon scattering rates $(1/\tau_{4,\lambda}^0)$ are calculated based on 4$^{th}$-FC as [49]

$$\frac{1}{\tau_{4,\lambda}^0} = \sum_{\lambda'\lambda''\lambda'''} \left\{ \frac{1}{6}\frac{n_{\lambda'}^0 n_{\lambda''}^0 n_{\lambda'''}^0}{n_\lambda^0}\mathcal{L}_{--} + \frac{1}{2}\frac{(1+n_{\lambda'}^0)n_{\lambda''}^0 n_{\lambda'''}^0}{n_\lambda^0}\mathcal{L}_{+-} + \frac{1}{2}\frac{(1+n_{\lambda'}^0)(1+n_{\lambda''}^0)n_{\lambda'''}^0}{n_\lambda^0}\mathcal{L}_{++} \right\} \quad (9)$$

Here $\lambda' = (\mathbf{q}', j')$, $\lambda'' = (\mathbf{q}'', j'')$, and $\lambda''' = (\mathbf{q}''', j''')$ go over all the phonon modes except for $\lambda = (\mathbf{q}, j)$. The formalisms of $\mathcal{L}_{--}$, $\mathcal{L}_{+-}$ and $\mathcal{L}_{++}$ can be found in Ref. [51]. Phonon natural isotope scattering is included. The exact solution of BTE is solved using the Fourphonon package [52] to obtain the phonon thermal conductivity:

$$\kappa_{ph}^{\alpha\beta} = \frac{\hbar^2}{k_B T^2 V_c N_\mathbf{q}} \sum_\mathbf{q}^{N_\mathbf{q}} \sum_j^{3n_b} v_\lambda^\alpha v_\lambda^\beta \omega_\lambda^2 n_\lambda(n_\lambda + 1)\tau_\lambda \quad (10)$$

Here $\lambda = (\mathbf{q}, j)$ goes over all phonon modes. $\tau_\lambda$ is the phonon lifetime after iteration.

The wavelike diffuson contribution is calculated by Wigner's formalism as [19,20]

$$\kappa_D^{\alpha\beta} = \frac{\hbar^2}{k_B T^2 V_c N_\mathbf{q}} \sum_\mathbf{q}^{N_\mathbf{q}} \sum_{j\neq j'}^{3n_b,3n_b} v_{\mathbf{q},jj'}^\alpha v_{\mathbf{q},j'j}^\beta \frac{\omega_\lambda + \omega_{\lambda'}}{2} \frac{\omega_\lambda n_\lambda(n_\lambda+1)+\omega_{\lambda'} n_{\lambda'}(n_{\lambda'}+1)}{4(\omega_{\lambda'}-\omega_\lambda)^2+(\tau_\lambda^{-1}+\tau_{\lambda'}^{-1})^2}(\tau_\lambda^{-1}+\tau_{\lambda'}^{-1}), \quad (11)$$

Here, $\lambda$ and $\lambda'$ represent the modes $(\mathbf{q},j)$ and $(\mathbf{q},j')$ with the same wavevector but different branches. $v_{\mathbf{q},jj'}^\alpha$ is the off-diagonal velocity between modes $(\mathbf{q},j)$ and $(\mathbf{q},j')$. The total thermal conductivity is obtained by adding both phonon contribution and diffuson contribution. The $\mathbf{q}$ mesh convergence test is carried out and discussed in detail.

## III. RESULTS

### A. Thermal expansion coefficient

The obtained GS lattice constant of $La_2Zr_2O_7$ in this work is 10.77 Å, in good agreement with experimental data, which is around 10.80 Å [11,53,54]. The calculated bulk modulus ($B$) is 168 GPa, which matches the literature value of 165-179 GPa [54,55]. Using the Grüneisen parameters ($\gamma_\lambda$) derived from two different methods (QHA and 3$^{rd}$-FC), and two different $B$ values (168 and 179 GPa), we predict the thermal expansion coefficient using Eq. (1) and temperature-dependent lattice constant using Eq. (2). The predicted lattice constant change relative to room temperature ($a(T)/a_{300K}$) is shown in Fig. 2 (a). It is seen that the three curves predicted using different $\gamma_\lambda$ and $B$ match with each other in general with a small deviation at high temperatures only. Using $\gamma_\lambda$ from 3$^{rd}$-FC overestimates the thermal expansion. Using a larger $B$ value underestimates the thermal expansion. Note that both $\gamma_\lambda$ and $B$ should be temperature-dependent, and both usually decrease with temperature [26]. Our early work has demonstrated that the introduction of temperature-dependent $B$ and FC can result in a better agreement with experimental data at elevated temperatures for ZrC [25]. Here, we take the ground-state $\gamma_\lambda$ and $B$ since the prediction has already matched well with experimental data [56–58] as shown in Fig. 2 (a). The thermal expansion coefficient also agrees well with experimental data as shown in Fig. 2 (b). The slight mismatch at low temperature might be caused by the experimental uncertainty due to the lack of data points at low temperatures.

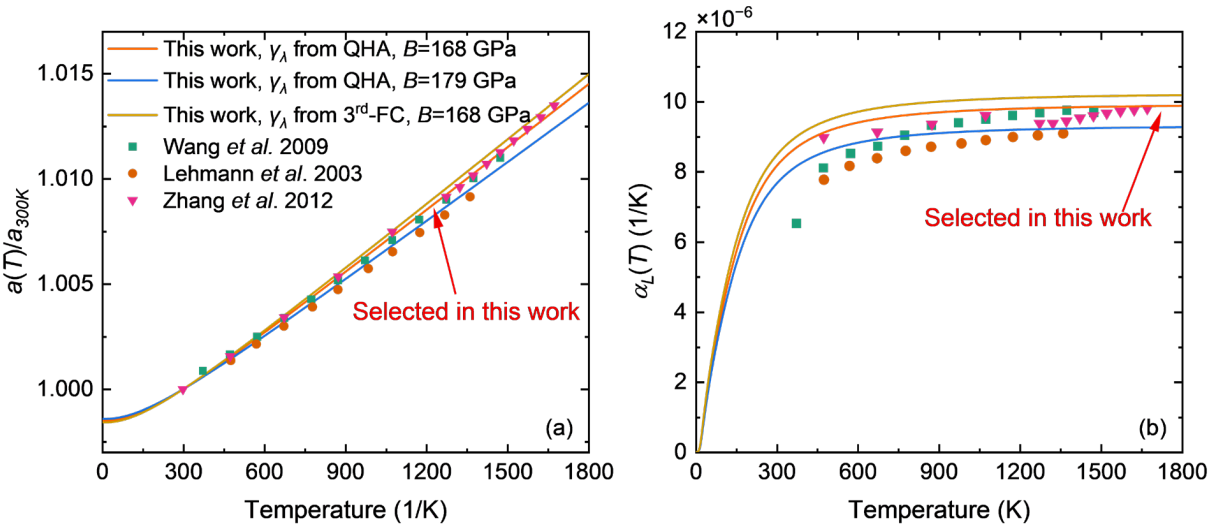

FIG. 2. (a) Temperature-dependent relative lattice constant ($a(T)/a_{300K}$) of La$_2$Zr$_2$O$_7$ predicted in this work compared to the experimental data. (b) Linear thermal expansion coefficient ($\alpha_L(T)$). Experimental data: Lehmann *et al*. 2003: [57], Wang *et al*. 2009: [58], and Zhang *et al*. 2012: [56].

**B. Machine learning interatomic potentials**

The MTP is trained and tested. As shown in Fig. 3(a), the training and testing error decreases with increasing MTP level and converges at the level of 22. Continuing to increase the level does not further increase the accuracy but increases the computational cost. Therefore, we pick level 22 for the potential training. At this level, the relative force errors for training and testing are less than 5% and 6%, respectively. To further test the accuracy, we calculate the phonon dispersion using the MLIP and find a good agreement with that obtained from DFT, as shown in Fig. 3(b). This indicates a high quality of our potential. The RMSE of energy and forces for snapshots in the testing database are 3.01 meV/atom and 0.081 eV/Å, as shown in Fig. 3(c,d), respectively. Both are relatively low, suggesting that the potential is sufficiently accurate.

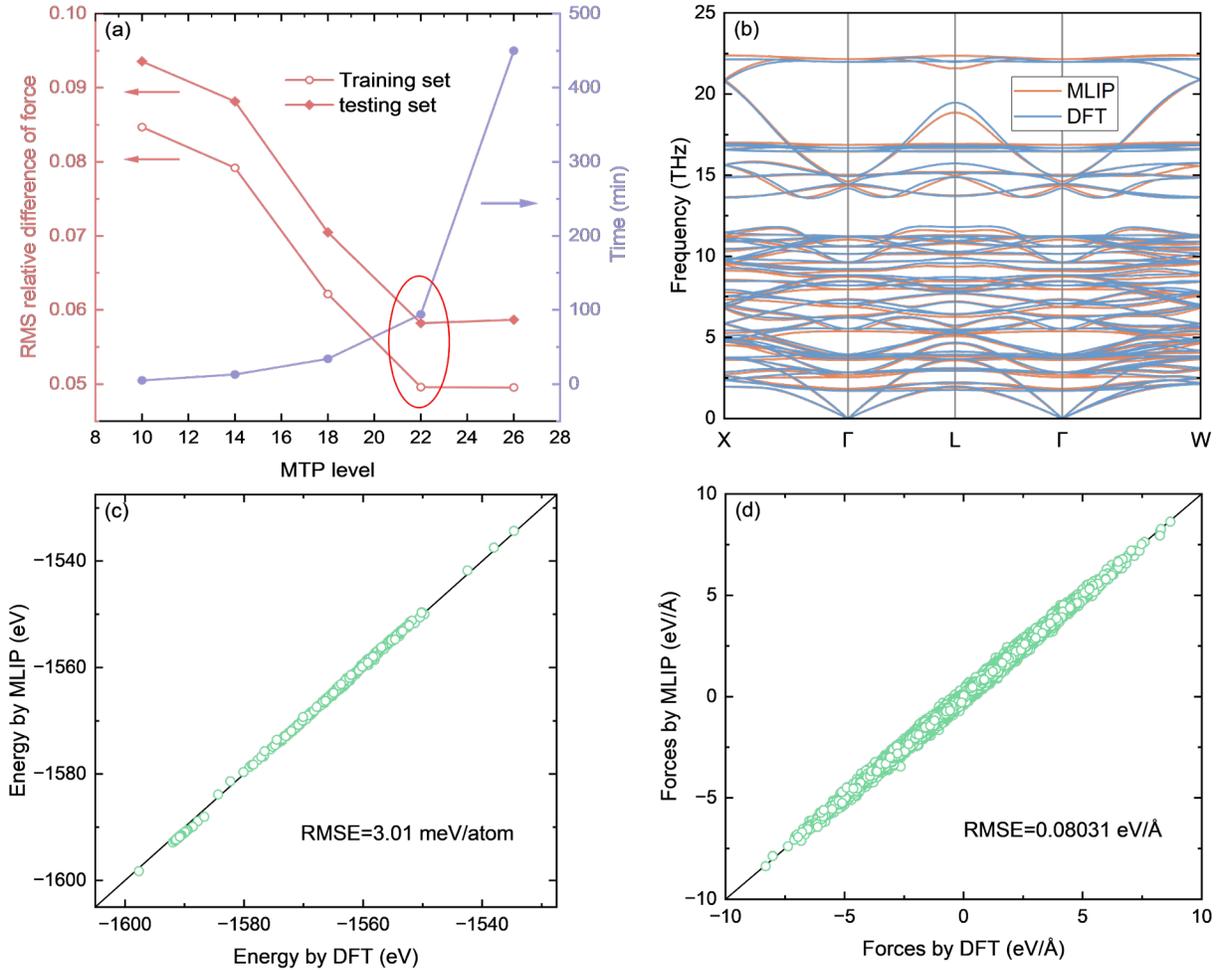

FIG. 3. (a) MTP level test. (b) Phonon dispersion calculated by MTP and DFT. (c) Energy and (d) forces predicted for configurations in the testing database from MTP and DFT.

**C. Thermal conductivity from GKMD**

With the MLIP, we perform MD simulations and extract thermal conductivities of $La_2Zr_2O_7$ at different temperatures using the GK formula. Fig. 4(a) showcases two examples of the smoothed integrated heat flux auto-correlation function, which presents well-converged values. This indicates that 50 ps is sufficient to capture the dissipation of heat current fluctuation. The domain size effect is shown in Fig. 4(b), which shows a good convergence at 4×4×4 supercell (5632 atoms) at room temperature. High temperature size convergence should be even faster as phonon lifetime decreases. This observation aligns with the finding in Ref. [59], which states that the supercell size convergence occurs earlier than **q**-mesh size convergence. Another apparent example is Si, whose GKMD converges at 8×8×8 supercell (4096 atoms) [59,60], while its BTE does not converge till 30×30×30 **q**-mesh [33,52]. Therefore, all the following GKMD results are based on the 4×4×4 supercell.

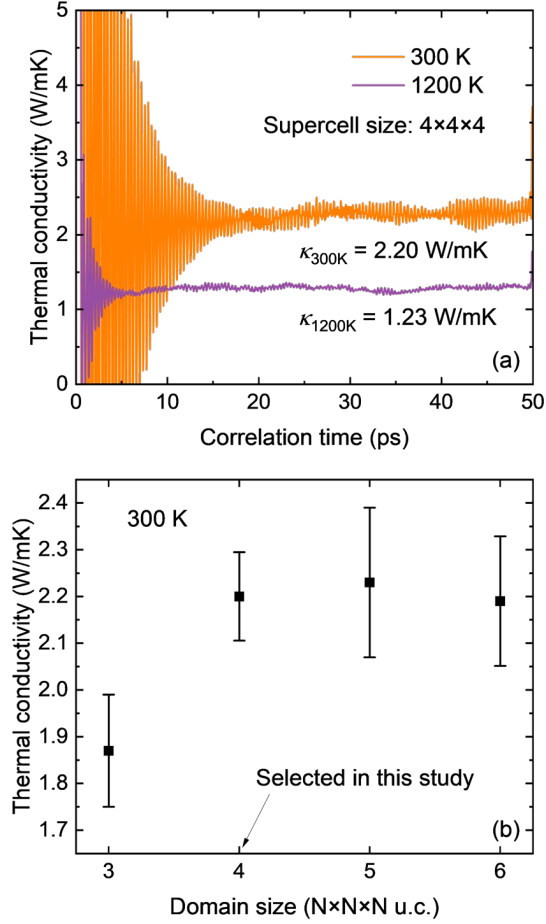

FIG. 4. (a) Thermal conductivity of $La_2Zr_2O_7$ obtained using GKMD at 300 K and 1200 K as a function of correlation time. (b) Thermal conductivity of $La_2Zr_2O_7$ obtained from GKMD at 300 K as a function of supercell size.

The obtained GK thermal conductivity of $La_2Zr_2O_7$ from 300 to 1800 K is shown in Fig. 5(a), which shows a reasonably good agreement with experimental data [6,8–12]. This agreement demonstrates the good accuracy of our MLIPs and GKMD simulations. Importantly, the GKMD successfully predicts the flat trend at ultra-high temperatures, indicating that the flat trend does not originate from some unknown mechanisms outside the GKMD. By fitting the thermal conductivity to a power law, we find that it is approximately $\sim T^{-0.40}$, being much flatter than $T^{-1}$.

It is noted that the experimental data taken from different literature papers slightly diverge from each other and also from GKMD. Some of them flatten while others increase with high temperatures, making the comparison between our theoretical prediction and the experimental data less convincing. To figure out the reason for the divergence of experimental data, we dig into the literature papers and find that all the experimental thermal conductivities ($\kappa_{\exp}$) are obtained by taking the product of thermal diffusivity ($D_{th}$), density ($\rho$), and specific heat capacity ($c_p$), i.e, $\kappa_{\exp} = \rho c_p D_{th}$. The difference of thermal conductivity mainly comes from the use of different specific heat $c_p$, as shown in Fig. 5(b). The experimental $c_p$ shows a diverging trend at high temperature, which is not consistent with lattice theory in solid state physics. In theory, the heat capacity per atom in any solid should increase with temperature and eventually converge at $3k_B/2$ as the kinetic energy of each atom converges at $3k_BT/2$. As seen in Fig. 5(b), the theoretical heat capacity converges after 1200 K. The diverging heat capacity obtained in experiments may be due to the formation of defects or a mistake in the measurement. It is noted that the heat capacity measurement should subtract the background data, which was often omitted in experiments in earlier years. To exclude the impact of the usage of different (or even wrong) $c_p$ in different papers, it is fair to compare the thermal diffusivity $D_{th}$ instead of thermal conductivity. As seen in Fig. 5(c), the experimentally measured $D_{th}$ matches well with our theoretical diffusivity obtained by $\kappa_{GK}/\rho c$. After excluding the impact of $c_p$, the experimental data all showed a flat trend instead of an increasing trend with temperature. In summary, our GKMD well predicts the absolute value and flat thermal diffusivity (thermal conductivity) of $La_2Zr_2O_7$ at ultra-high temperatures.

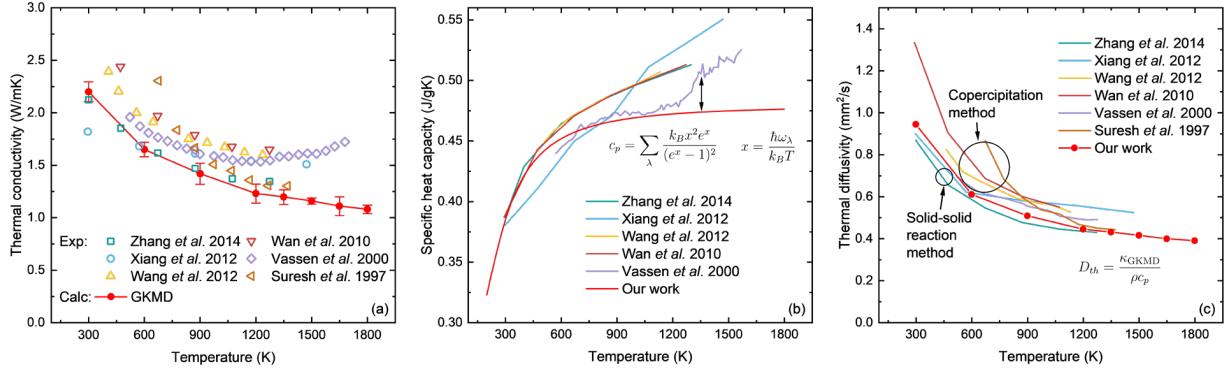

FIG. 5. (a) Thermal conductivity of La$_2$Zr$_2$O$_7$ using GKMD with MTP. Experimental data and Wigner formalism results are provided for reference. (b) Specific heat capacity and (c) thermal diffusivity of La$_2$Zr$_2$O$_7$. References of experimental data: Zhang *et al*. [8], Xiang *et al*. [10], Wang *et al*. [11], Wan *et al*. [12], Vassen *et al*. [6], and Suresh *et al*. [9].

### D. Thermal conductivity from SED-BTE (phonon particle)

In this section, we predict the thermal conductivity of La$_2$Zr$_2$O$_7$ by BTE using the phonon lifetime obtained from SED. Taking the reduced **q** point at (1/3, 0, 0) as an example, we show the SED as a function of frequency at 300 K and 1200 K in Fig. 6(a,b). At 300 K, each phonon branch at this given **q** point is a distinct peak. By fitting each peak to a Lorentzian function, the position of each peak tells the frequency of the phonon branch, and the linewidth gives the phonon scattering rates. The fitting results of an example phonon mode at a randomly selected **q** point (reduced **q** = [1/3, 0, 0]) at 300, 750, and 1200 K are shown in Fig. 6(c-e). With increasing temperature, the peaks are broadened and eventually overlap with each other. To distinguish the peaks from each other, eigenvectors are used to obtain the SED of each branch separately. The SED results at 1800 K is not shown here for the eigenvectors at such high temperature are no longer working. This is a drawback of the SED to study the thermal conductivity of materials at ultra-high temperatures.

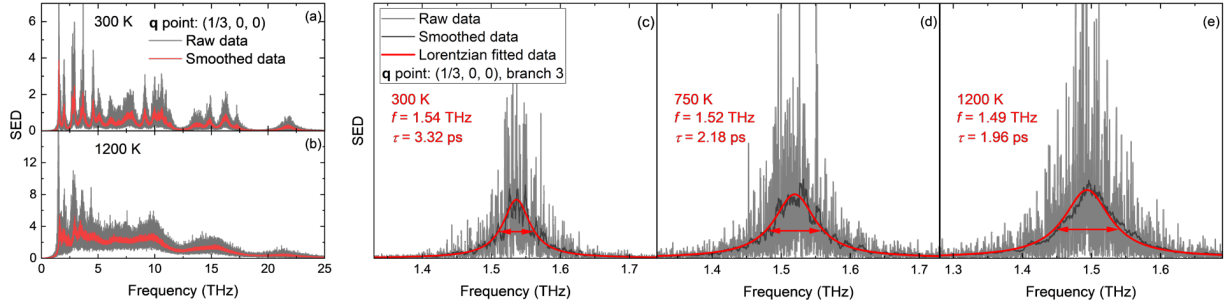

FIG. 6. Phonon total SED data for q point (1/3, 0, 0) at (a) 300 K and (b) 1200 K. SED data for a single phonon mode, i.e., the third branch of q point (1/3, 0, 0), at (c) 300 K, (d) 750 K and (e) 1200 K.

It might be perceived that the phonon lifetime obtained from SED depends on the MD simulation size. This is because the MD domain size determines how many phonon modes can be resolved, e.g., a $N{\times}N{\times}N$ primitive cell can only resolve $N{\times}N{\times}N$ **q** points. Larger domain size resolves more phonon modes, and phonons can more easily find other modes to scatter. However, we find that the obtained lifetimes from SED of a 4×4×4 cubic cell are the same as those obtained from a 6×6×6 cubic cell (shown in Fig. S1 in the Supplemental Material [61]). Therefore, the size effect of the SED is eliminated.

The obtained phonon scattering rates from SED are shown in Fig. 7(a,b) for 300 and 1200 K, respectively. Note that these scattering rates naturally include the impacts of all higher-order phonon-scattering, phonon renormalization, and phonon scattering cross-section softening effects at finite temperatures. In comparison, we also plot the 3ph and 4ph scattering rates calculated by using the GS force constants. It is seen that 4ph scattering is negligible at 300 K but non-negligible at 1200 K. We also note that the GS 3ph scattering rates match well with SED at both temperatures. This is likely a coincidence due to self-cancelling effect, i.e., the finite-temperature corrections to harmonic and anharmonic force constants reduce scattering rates and cancel out the addition of 4ph scattering. This coincidence is seen in many other materials [22,62] but not all materials [24–26,63].

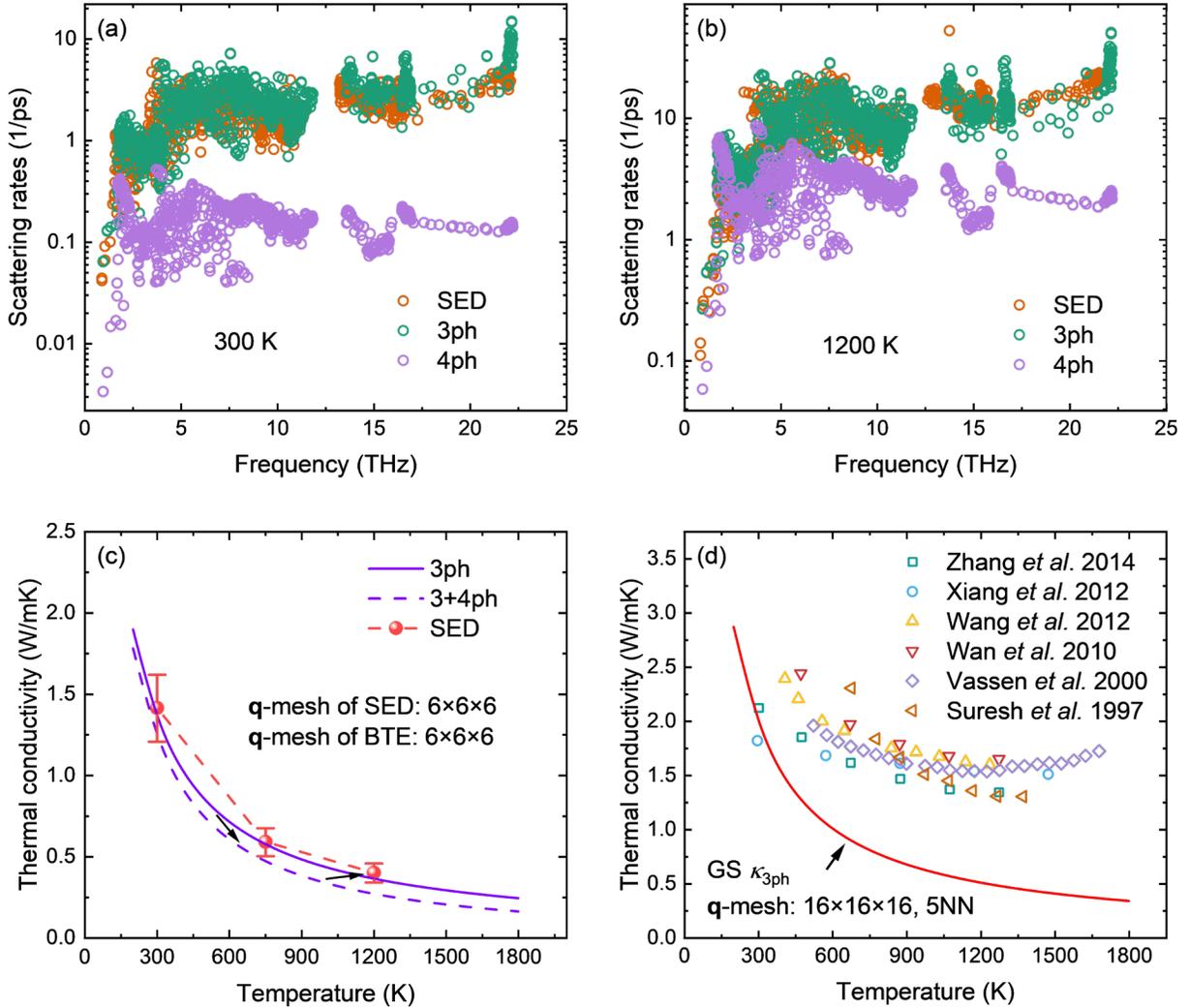

FIG. 7. Phonon scattering rates extracted using SED analysis and BTE (3ph and 4ph) at (a) 300 K and (b) 1200 K. The q-mesh sizes are all 6×6×6. (c) Thermal conductivity of $La_2Zr_2O_7$ within the framework of BTE using the phonon lifetimes obtained from SED analysis and BTE (3ph and +4ph). (d) The comparison of thermal conductivity between BTE (3ph) and experimental data.

Since 3ph thermal conductivity agrees with SED thermal conductivity, this figure can also be seen as the compassion between SED and experimental data. References of experimental data: Zhang et al. [8], Xiang et al. [10], Wang et al. [11], Wan et al. [12], Vassen et al. [6], and Suresh et al. [9].

With the scattering rates, the thermal conductivity of $La_2Zr_2O_7$ within the phonon particle framework is calculated by BTE, as shown in Fig. 7(c). To consistently compare the results from temperature-dependent SED and ground-state 3+4ph scattering, we pick the same **q**-mesh of 6×6×6 in Fig. 7(c). 4ph is not important at room temperature but becomes significant at ultra-high temperatures, which can result in a 35% reduction in thermal conductivity compared with 3ph. But this reduction is recovered by the temperature corrections to the harmonic and anharmonic force constants. As a result, the thermal conductivity from SED matches well with that from GS 3ph BTE. To compare with experimental data, we calculate the BTE phonon thermal conductivity converged with respect to the **q**-mesh density. Since SED is limited by the MD simulation domain size, we use the GS 3ph scattering to represent the SED results, since they are similar. As seen in Fig. 7(d), the BTE, even with temperature corrections, cannot predict the flat thermal conductivity $La_2Zr_2O_7$ at high temperatures.

### E. Thermal conductivity from Wigner formalism (phonon + diffuson)

The calculated temperature-dependent thermal conductivity from Wigner formalism is shown in Fig. 8(a). The total thermal conductivity, which consists of phonon and diffuson contributions, aligns with experimental data across a broad temperature range. Additionally, it exhibits agreement with $\kappa_{GK}$ throughout the entire temperature range, as illustrated in Fig. S2 in the Supplemental Material [61], despite operating on different principles: all-order phonon scatterings and all the finite-temperature corrections are inherently considered in GKMD, while up to three-phonon scattering and GS phonon properties are accounted for in Wigner formalism. Based on the aforementioned discussion in SED part, the higher-order phonon scattering appears to counterbalance the effect of finite-temperature corrections. Therefore, this indicates that diffuson is the primary cause of the flat thermal conductivity of $La_2Zr_2O_7$ at ultra-high temperatures, which contributes approximately 70% to thermal conductivity. The fitting of the calculated data gives scaling law of $\sim T^{-0.97}$ and $\sim T^{0.43}$ for $\kappa_{ph}$ and $\kappa_D$, respectively.

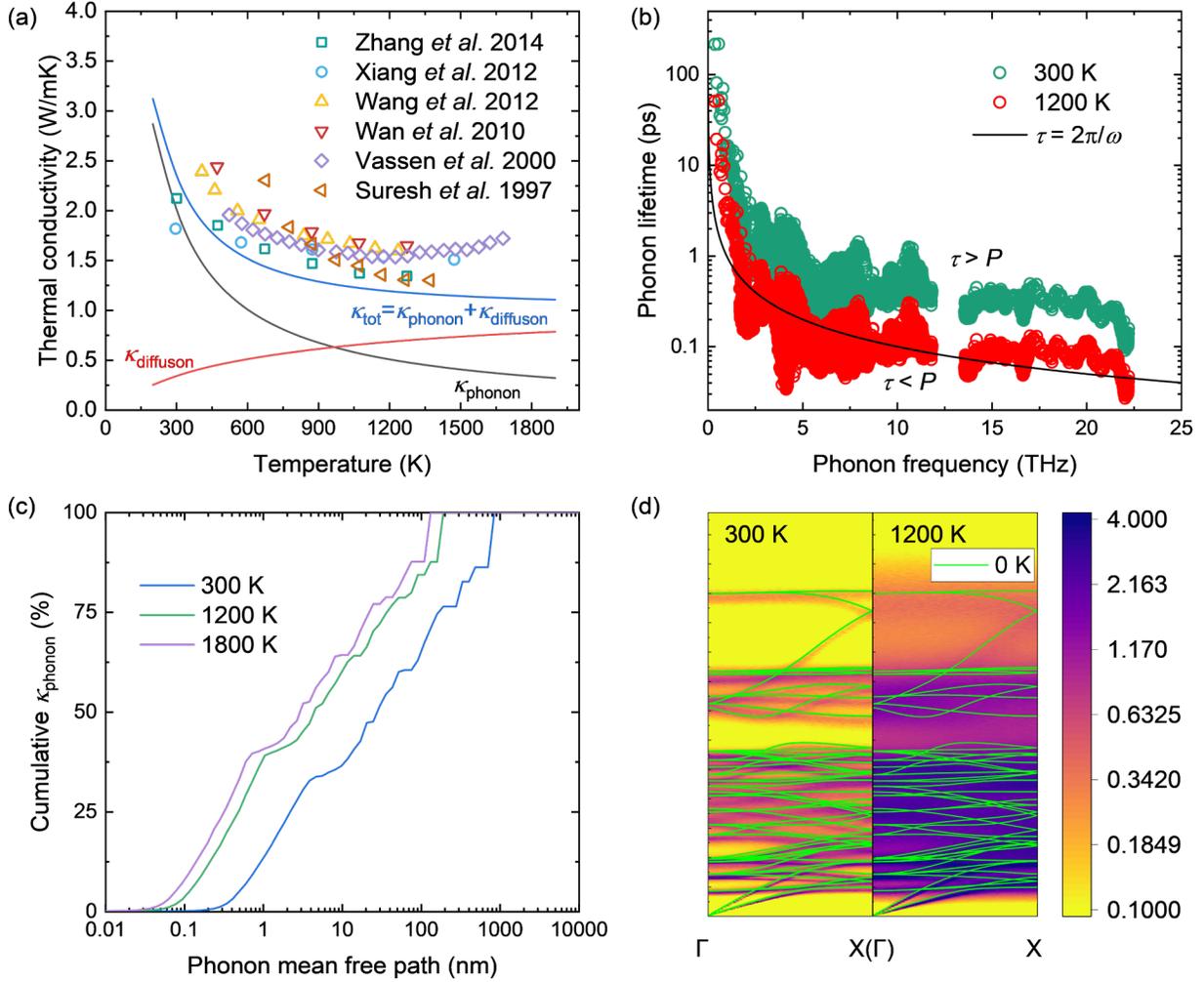

FIG. 8. (a) Temperature-dependent thermal conductivity of $La_2Zr_2O_7$ obtained using Wigner formalism. References of experimental data: Zhang et al. [8], Xiang et al. [10], Wang et al. [11], Wan et al. [12], Vassen et al. [6], and Suresh et al. [9]. (b) Phonon lifetime at 300 and 1200 K by solving BTE with 16×16×16 **q**-mesh for 3ph. (c) Percentage cumulative phonon thermal conductivity with respect to mean free path at 300, 1200, and 1800 K. (d) Phonon dispersion obtained from SED analysis at 300 K and 1200 K, compared to the ground state dispersion.

To gain more insights regarding the diffuson contribution, we show the lifetimes of all the phonons at 300 and 1200 K for 3ph in Fig. 8(b). The black curves indicate the limit that phonon lifetime ($\tau$) equals period ($2\pi/\omega$). At a higher temperature, more phonon modes drop below this curve and become more diffuson-like. Surprisingly, even at ultra-high temperatures (e.g., T>1200 K), the phonon mean free path can still spread over a broad range from 0.1 to 200 nm, as seen in Fig. 8(c). This indicates that grain boundary scattering can still limit phonon transport if the grain size is in the order of 100 nm. The phonon dispersions at 300 and 1200 K extracted using SED analysis are shown in Fig. 8(d). We find that higher temperatures broaden the phonon linewidth, making different branches overlap with each other. This provides the channels for vibrations to tunnel between close branches, namely, inter-band diffuson transport. This result from SED agrees with that calculated from the Wigner formalism. In addition, the softening effect caused by finite-temperature corrections can also be clearly seen in Fig. 8(d).

The **q**-mesh convergence test is studied within the Wigner formalism framework, as illustrated in Fig. 9(a). Similar to phonon, diffuson thermal conductivity also increases with the density of **q**-mesh. It might be perceived that the impact of **q**-mesh on thermal conductivity may come from the impact on phonon lifetimes since it is easier for phonons to find others to scatter in a denser **q**-mesh. However, interestingly, as shown in Fig. 9(b), we find that the density of **q**-mesh has no effect on the lifetime. The phonon modes in a 4×4×4 **q**-mesh exhibit nearly identical lifetimes as those of a 16×16×16 **q**-mesh. This indicates that the impact of **q**-mesh on thermal conductivity arises from the sampling resolution of the Brillouin zone rather than phonon lifetime.

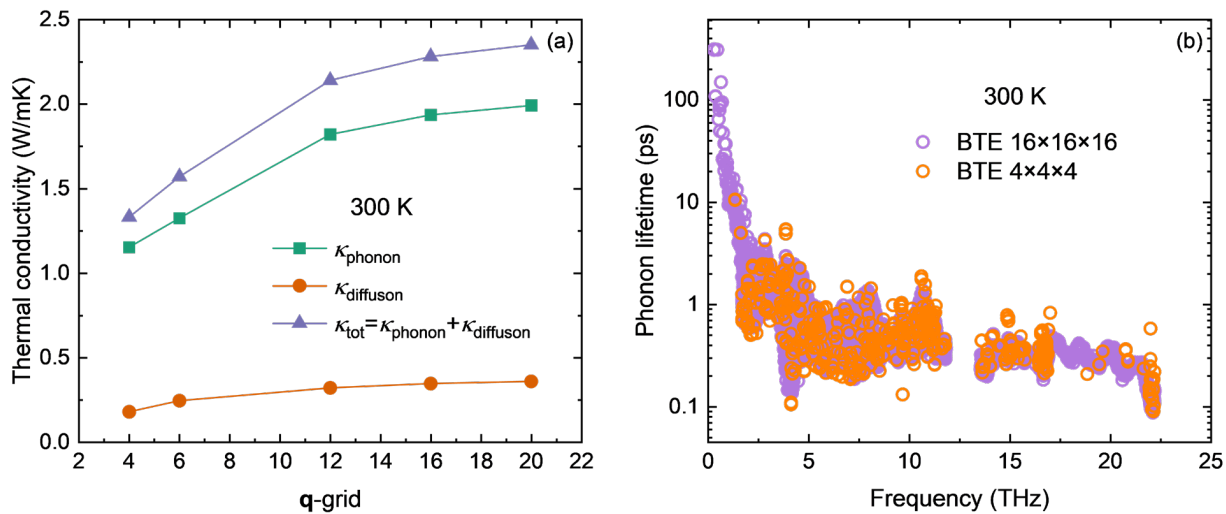

FIG. 9. (a) thermal conductivity obtained using Wigner formalism as a function of **q**-mesh density. (b) Lifetime of phonons in the framework of BTE with different **q**-mesh densities.

The impact of non-analytical correction (NAC) on thermal conductivity is also studied. As shown in Fig. 10(a), the inclusion of NAC increases the phonon thermal conductivity and decreases the diffuson contribution slightly, resulting in an increase in the total thermal conductivity. To find out the reasons, we show the phonon lifetimes, velocities, and dispersions calculated with and without NAC, as shown in Fig. 10(b-d). With NAC, phonon lifetimes do not change significantly, but the velocities are decreased for certain modes since NAC can cause TO-LO splitting, which flattens some branches in phonon dispersion. Since charges are not considered MLIP, MD is not able to account for the NAC. Therefore, the results displayed above do not consider NAC. The inclusion of NAC can make the thermal conductivity slightly more flattened at high temperatures, and the difference caused by NAC becomes smaller at higher temperatures.

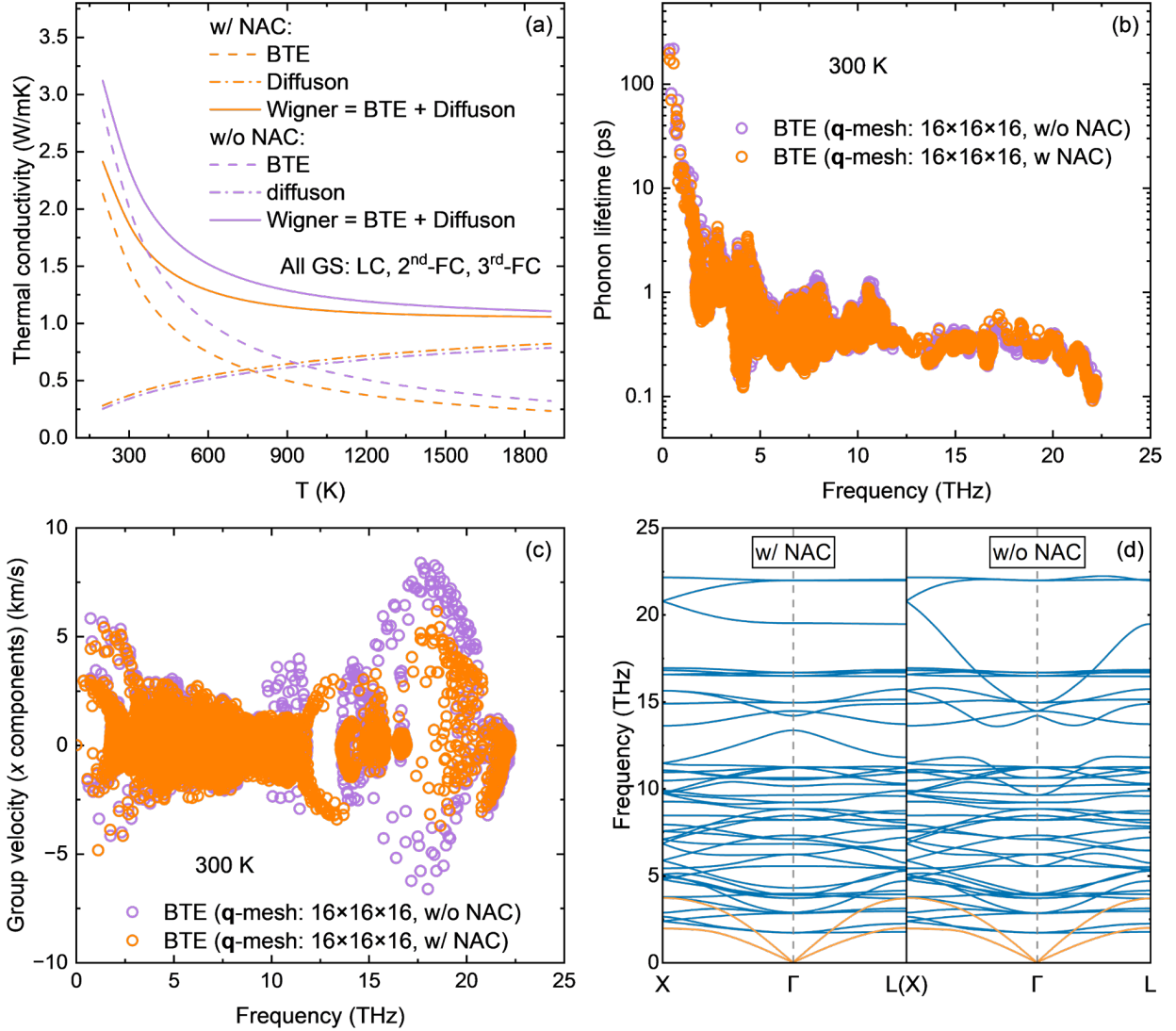

FIG. 10. The impact of non-analytical correction on (a) temperature-dependent thermal conductivity, (b) phonon lifetime at 300 K, (c) group velocities (*x* component) at 300 K, and (d) phonon dispersion.

**F. Radiation contribution to thermal conductivity**

It has been perceived that radiation contributes to the thermal conductivity of $La_2Zr_2O_7$ at high temperatures in the earlier experimental works based on the flat/increasing thermal conductivity trend [64,65]. However, no evidence has been given so far. Here, we rigorously calculate the radiation contribution to thermal conductivity as a function of temperature fully from the first principles. Analogous to phonons, photons can also conduct thermal energy within materials. Phonon mean free paths are limited by creation and annihilation processes (i.e., scattering events), and similarly, photon mean free paths are governed by absorption and re-emission events. When a material's size significantly exceeds its phonon mean free path, phonon transport becomes diffusive. Similarly, when a material's size significantly exceeds its photon mean free path, photon transports diffusively, and the material becomes optically thick. In the diffusive limit, the radiation contribution to the thermal conductivity of a material ($\kappa_{rad}$) can be calculated by the Rosseland model $\kappa_{rad} = \frac{16}{3\beta} n^2 \sigma_{SB} T^3$, where $\beta$ is the extinction coefficient, $n$ is refractive index, and $\sigma_{SB}$ is the Stefan-Boltzmann constant. $\beta$ and $n$ are temperature dependent. The details of the method are given in Ref. [63]. In this method, the input is primarily the absorption coefficient and dielectric function, which can be calculated from the Lorentz oscillator model with phonon linewidths calculated from first principles.

The results are shown in Fig. 11. The real part of the refractive index at the high-frequency limit is 2.22, which matches the experimental data of 2.17 [66]. The imaginary part of the refractive index is related to the extinction coefficient, which measures the attenuation of radiative waves inside the medium and is inversely correlated to photon MFP or penetration depth. The higher the extinction coefficient, the lower the radiation thermal conductivity. At room temperature, photons have a penetration depth of around 35 μm at the wavelength of 10 μm (the dominant thermal radiation wavelength at room temperature). At high temperatures (e.g., 1800 K), the penetration depth decreases to 10 μm at the dominant thermal radiation wavelength of 1.6 μm. This is understandable as there are more phonons to scatter photons at higher temperatures. $\kappa_{rad}$ increases with temperature by an approximate power law of $\sim T^{2.01}$, fitted from the calculated data. The extinction coefficient $\beta$ increases approximately linearly with temperature. At the maximum temperature, 1800 K, $\kappa_{rad}$ is found to be only 0.074 W/mK, which is much smaller compared to phonon and diffuson thermal conductivity. This provides a critical revisit to the preliminary assumption that radiation is important at high temperatures for $La_2Zr_2O_7$. This result is consistent with the finding from the previous sections that the flat trend of thermal conductivity of $La_2Zr_2O_7$ can be well explained by the lattice vibrations (e.g., phonon+diffuson) without the need of radiation.

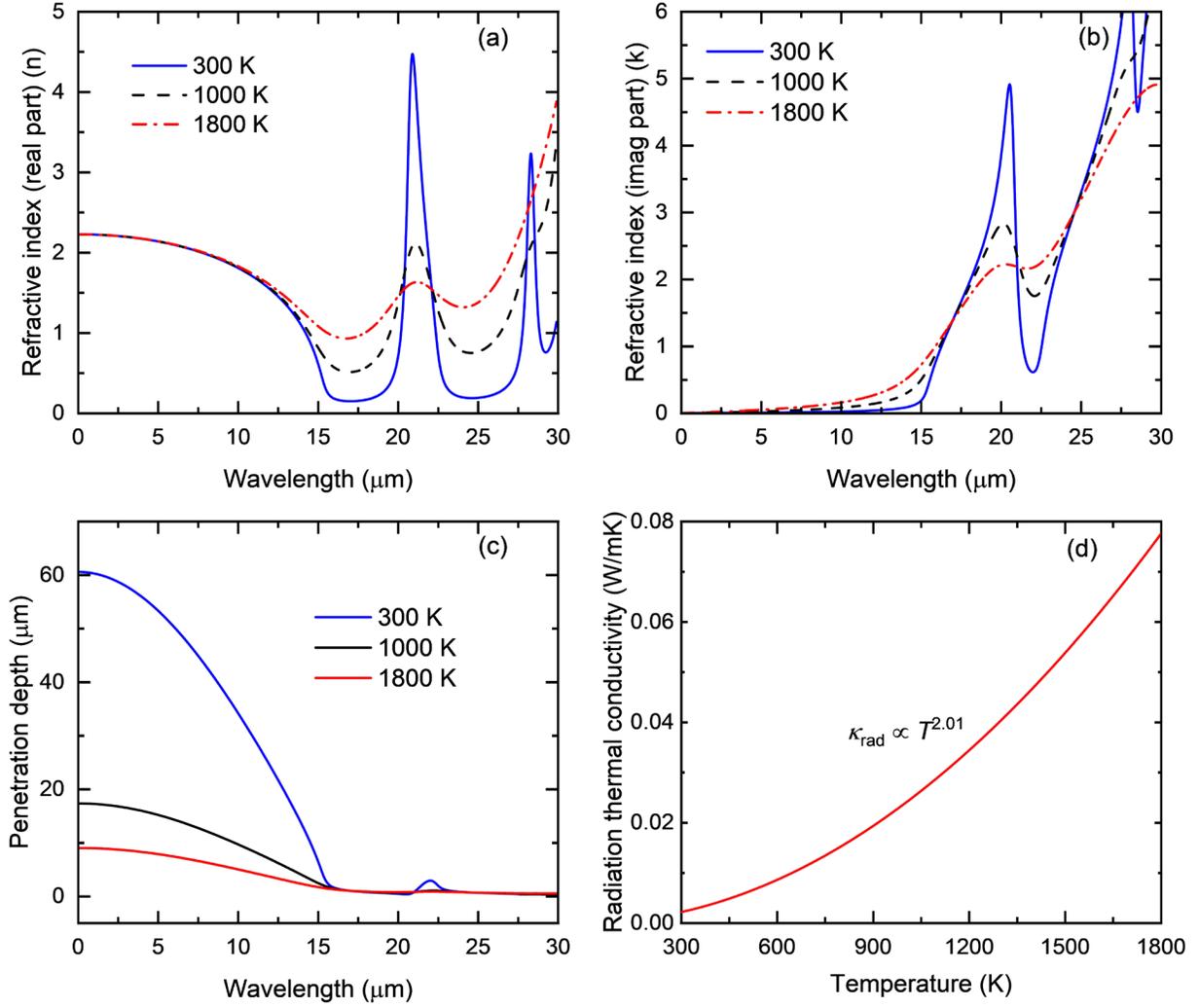

FIG. 11. Spectral radiative thermal properties of $La_2Zr_2O_7$ calculated from first principles. Ground state force constants are used in the calculations. (a) Real part, (b) imaginary part of the refractive index calculated from the Lorentz oscillator model using the phonon linewidths calculated from the 3ph and 4ph scattering. (c) Penetration depth of photons. (d) Radiation thermal conductivity as a function of temperature. $T^2$ is a fitted power law using the calculated $\kappa_{rad}$ data.

## IV. CONCLUSIONS

We have predicted and compared the thermal conductivity of $La_2Zr_2O_7$ using three distinct first-principles methods, namely GKMD with machine learning potentials, Peierls BTE with finite-temperature corrected phonon lifetimes extracted using SED, and Wigner formalism. The following conclusions can be drawn: (i) GKMD with machine learning potential is able to predict the absolute value and flat temperature dependency of thermal diffusivity (thermal conductivity) of $La_2Zr_2O_7$ from room to ultra-high temperatures. (ii) 4ph is important at ultra-high temperatures, which can lead to a 35% deduction in thermal conductivity of 3ph. (iii) Finite-temperature corrections can cancel out 4ph and bring up the 3+4ph thermal conductivity back to the ground state 3ph level for $La_2Zr_2O_7$. (iv) Finite-temperature corrections within Peierls BTE framework cannot predict the flat trend of thermal conductivity of $La_2Zr_2O_7$ at ultra-high temperatures. (v) Diffuson contribution is primarily responsible for the flat trend at ultra-high temperatures in the thermal conductivity of $La_2Zr_2O_7$. (vi) Radiation contribution to thermal conductivity of $La_2Zr_2O_7$ increases with $T^{2.01}$, but is still small compared to phonon and diffuson contributions even at ultra-high temperatures. (vii) GKMD does not have a significant size effect, e.g., thousands of atoms are sufficient. (viii) The **q** mesh density has a negligible effect on scattering rates but a pronounced impact on thermal conductivity due to the sampling resolution problem in the Brillouin zone. (ix) The scaling law of phonon, diffuson, radiation, and total thermal conductivity of $La_2Zr_2O_7$ is about $\sim T^{-0.97}$, $\sim T^{0.43}$, $\sim T^{2.01}$, and $\sim T^{-0.40}$, respectively. We hope this study clarifies thermal transport mechanisms in $La_2Zr_2O_7$ from room to ultra-high temperatures and provides inspiration for other similarly complex materials for ultra-high temperature applications.

**Data availability**

Source data are provided with this paper. All other data that support the plots within this paper are available from the corresponding authors on reasonable request.

**Code availability**

The codes used in this study are available from the corresponding authors upon request.

**Acknowledgments**

This work is supported by the National Science Foundation (NSF) (Award No. CBET 2212830). The computation used Center for High Performance Computing (CHPC) at the University of Utah and Bridges-2 at Pittsburgh Supercomputing Center through allocation PHY220002 from the Advanced Cyberinfrastructure Coordination Ecosystem: Services & Support (ACCESS) program, which is supported by NSF grants #2138259, #2138286, #2138307, #2137603, and #2138296.

**Author contributions**

H.Z. performed the simulations. J.T. performed the radiation calculations. T.F. conceived the idea and guided the project. H.Z. wrote the original manuscript. T.F. revised the manuscript.

**Competing interests**

The authors declare no competing interests.

**Additional information**

Correspondence and requests for materials should be addressed to T.F.